\documentclass{iau}
\usepackage{graphicx}
\usepackage{natbib} 
\title[Keenan et al.,~~Local Large-Scale Structure] %% give here short title %%
{Local Large-Scale Structure and The Assumption of Homogeneity}

\author[R. C. Keenan et al.]   %% give here short author list %%
{Ryan C. Keenan$^1$, Amy J. Barger$^{2,3,4}$, \and Lennox L. Cowie$^3$}
\affiliation{$^1$Academia Sinica Institute of Astronomy and Astrophysics\\ P.O. Box 23-141, Taipei 10617, Taiwan \\ email: {\tt rkeenan@asiaa.sinica.edu.tw} \\[\affilskip]
$^2$Dept. of Astronomy, University of Wisconsin-Madison\\ 475 N. Charter St., Madison, WI 53706, USA\\[\affilskip]
$^3$Dept. of Physics and Astronomy, University of Hawaii\\ 2505 Correa Rd., Honolulu, HI 96822, USA\\[\affilskip]
$^4$Institute for Astronomy, University of Hawaii, 2680 Woodlawn Dr., Honolulu, HI 96822, USA}

\pubyear{2014}
\volume{308}  %% insert here IAU Symposium No.
\pagerange{119--126}
\jname{The Zeldovich Universe}
\editors{R. van de Weygaert, S. Shandarin, E. Saar, \& J. Einasto}
\begin{document}

\maketitle

\begin{abstract}
Our recent estimates of galaxy counts and the luminosity density in the near-infrared \citep{Keen10a,Keen12} indicated that the local universe may be under-dense on radial scales of several hundred megaparsecs.  Such a large-scale local under-density could introduce significant biases in the measurement and interpretation of cosmological observables, such as the inferred effects of dark energy on the rate of expansion.  In \citet{Keen13}, we measured the $K-$band luminosity density as a function of distance from us  to test for such a local under-density.  We made this measurement over the redshift range $0.01 < z < 0.2$ (radial distances $D\sim 50-800$~$h_{70}^{-1}$~Mpc).  We found that the shape of the  $K-$band luminosity function is relatively constant as a function of distance and environment.   We derive a local ($z < 0.07, D < 300$~$h_{70}^{-1}$~Mpc) $K-$band luminosity density that agrees well with previously published studies.   At $z > 0.07$, we measure an increasing luminosity density that by $z\sim 0.1$ rises to a value of $\sim 1.5$ times higher than that measured locally.   This implies that the stellar mass density follows a similar trend.    Assuming that the underlying dark matter distribution is traced by this luminous matter, this suggests that the local mass density may be lower than the global mass density of the universe at an amplitude and on a scale that is sufficient to introduce significant biases into the measurement of basic cosmological observables.  At least one study has shown that an under-density of roughly this amplitude and scale could resolve the apparent tension between direct local measurements of the Hubble constant and those inferred by Planck team.   Other theoretical studies have concluded that such an under-density could account for what looks like an accelerating expansion, even when no dark energy is present. 
\end{abstract}

\firstsection % if your document starts with a section,
\section{Introduction}

The assumption that the universe is homogeneous on large scales is a fundamental pillar of our current concordance cosmology.  However, it is well known that very large-scale inhomogeneity exists in the form of sheets, voids, and filaments of matter .  As structure formation proceeds, gravity acts to pile more matter onto the over-dense regions, and simultaneously evacuate the under-dense regions.  Models have shown that this phenomenon implies that cosmological observables measured by an observer will vary significantly depending on where that observer is located with respect to these large-scale structures.

\section{Our Study and its Implications}
In two previous papers \citep{Keen10a, Keen12}, we investigated large-scale inhomogeneity in the local universe.  These studies were inspired, in part, by the fact that a number of recent cosmological modeling efforts have shown that the observational phenomena typically used to infer an accelerating expansion, namely the ``dimming" of type Ia supernovae, can be equally well fit by invoking a large local under-density instead of dark energy (e.g., \citealt{Alex09, Bole11a}).  Other models \citep{Marr13} have shown that, even in the context of a universe dominated by dark energy, a large local under-density could explain the apparent discrepancy between local measurements of the Hubble constant  and those inferred by the Planck team \citep{Ade13}.

In the study presented here, we investigated the $K-$band luminosity density (as a proxy for stellar mass density) as a function of distance from our position in the local universe.   To accomplish this we combined photometry from the UKIRT Infrared Deep Sky Large Area Survey (UKIDSS-LAS, \citealt{Lawr07}) and the Two Micron All Sky Survey Extended Source Catalog (2MASS-XSC, \citealt{Skru06}) with redshifts from the the Sloan Digital Sky Survey (SDSS, \citealt{York00}), the Two-degree Field Galaxy Redshift Survey (2dFGRS, \citealt{Coll01}), the Galaxy And Mass Assembly Survey (GAMA, \citealt{Driv11}) the Two Micron Redshift Survey (2MRS, \citealt{Erdo06}), and the Six Degree Field Galaxy Redshift Survey (6DFGS, \citealt{Jone09}). 

Key results from our study \citep{Keen13} are presented in the figures below.  In Figure 1a, we show our measurement of the $K-$band luminosity function using our compiled sample in three separate redshift ranges (with residuals showing the normalization offset between these).  In Figure 1b, we show the relative contribution to the total light as a function of absolute magnitude to demonstrate that we are resolving $\sim 75\%$ of the total light at $z < 0.2$.  In Figure 2a, we show a comparison of our estimate of the total luminosity density in different directions on the sky.  In Figure 2b, we show our results compared to other studies from the literature and models.  We conclude that the local universe appears under-dense on a scale and amplitude sufficient to introduce significant biases into local measurements of the expansion rate, and, according to some models, to cause what looks like an accelerating expansion even when no dark energy is present.  

 %*******************
% FIGURE 9
%*******************
\begin{figure}
\begin{center}
\includegraphics[width=120mm]{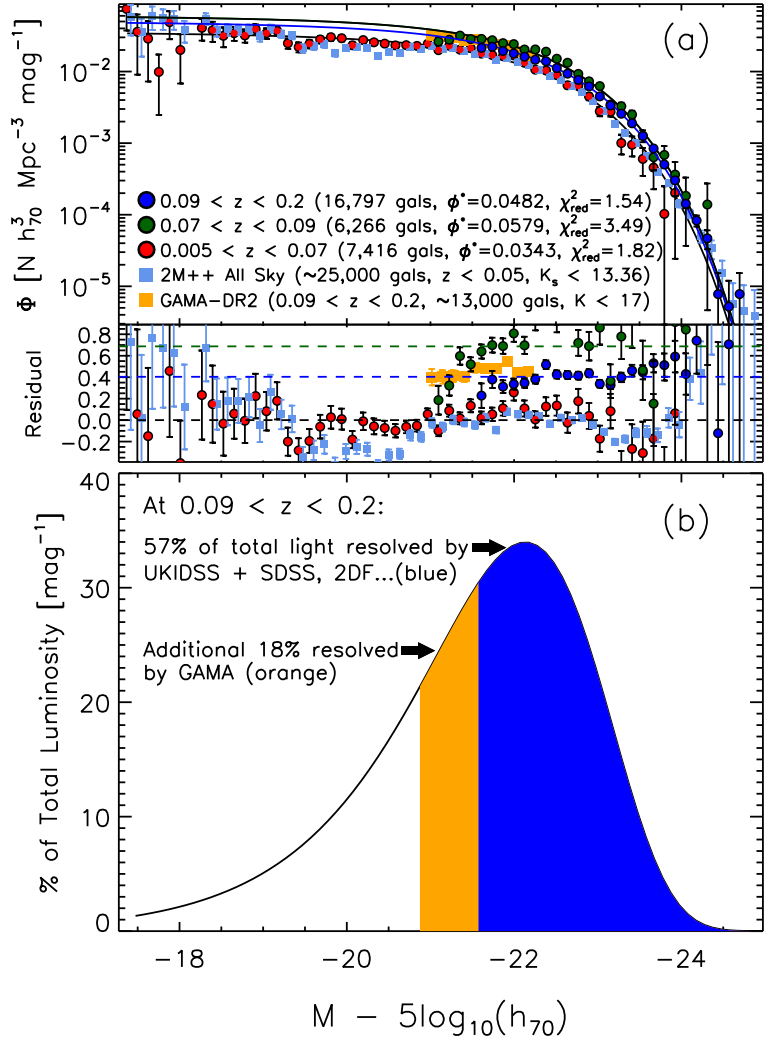}
\caption{\label{lovhi} (a) The UKIDSS-LAS $K-$band luminosity function split into three redshift ranges: $0.005<z<0.07$ (red), $0.07<z<0.09$ (green), and $0.09< z < 0.2$ (blue).  We separate the redshift range $0.07<z<0.09$ to demonstrate that the excess at $z > 0.09$ is not due to the Sloan Great Wall or the other over-densities we observe at higher declination in this redshift range.    Here we fit $\phi^*$ in each redshift range with $\alpha$ and $M^*$ fixed (as described in \citealt{Keen13}).  We list the $\chi^2_{\rm{red}}$ values for each redshift range in the figure, and note that relatively good fits can be obtained using fixed LF shape parameters and letting the normalization ($\phi^*$) vary as a function of redshift.   Residuals are shown relative to the low-redshift normalization.  Here we also include our own analysis of the 2M++ catalog (all sky, $K_{\rm{s,AB}}<13.36,~\sim 25,000$ galaxies).  We have fit the normalization of the LF derived from the 2M++ catalog (LF and residuals shown as light blue squares) with the same shape parameters as for the UKIDSS sample.   To extend the faint end of the $z>0.09$ LF, we use the GAMA redshift sample ($K_{\rm{AB}}<17$, shown as orange squares).  (b) The relative contribution to the total luminosity density (\% per magnitude) as a function of absolute magnitude.  The blue shaded region shows the range of magnitudes covered by the UKIDSS sample at $z>0.09$.  In orange we show the extra fraction of light resolved by extending the faint end of the LF with the GAMA sample.  This demonstrates that, at $z>0.09$, we are making a robust measurement of the peak of the luminosity density distribution (occurring at $M\sim M^*$) and resolving $\sim 75\%$ of the total light.}
\end{center}
\end{figure} 
%*******************

%*******************
% FIGURE 11
%*******************
\begin{figure}
\begin{center}
\includegraphics[width=120mm]{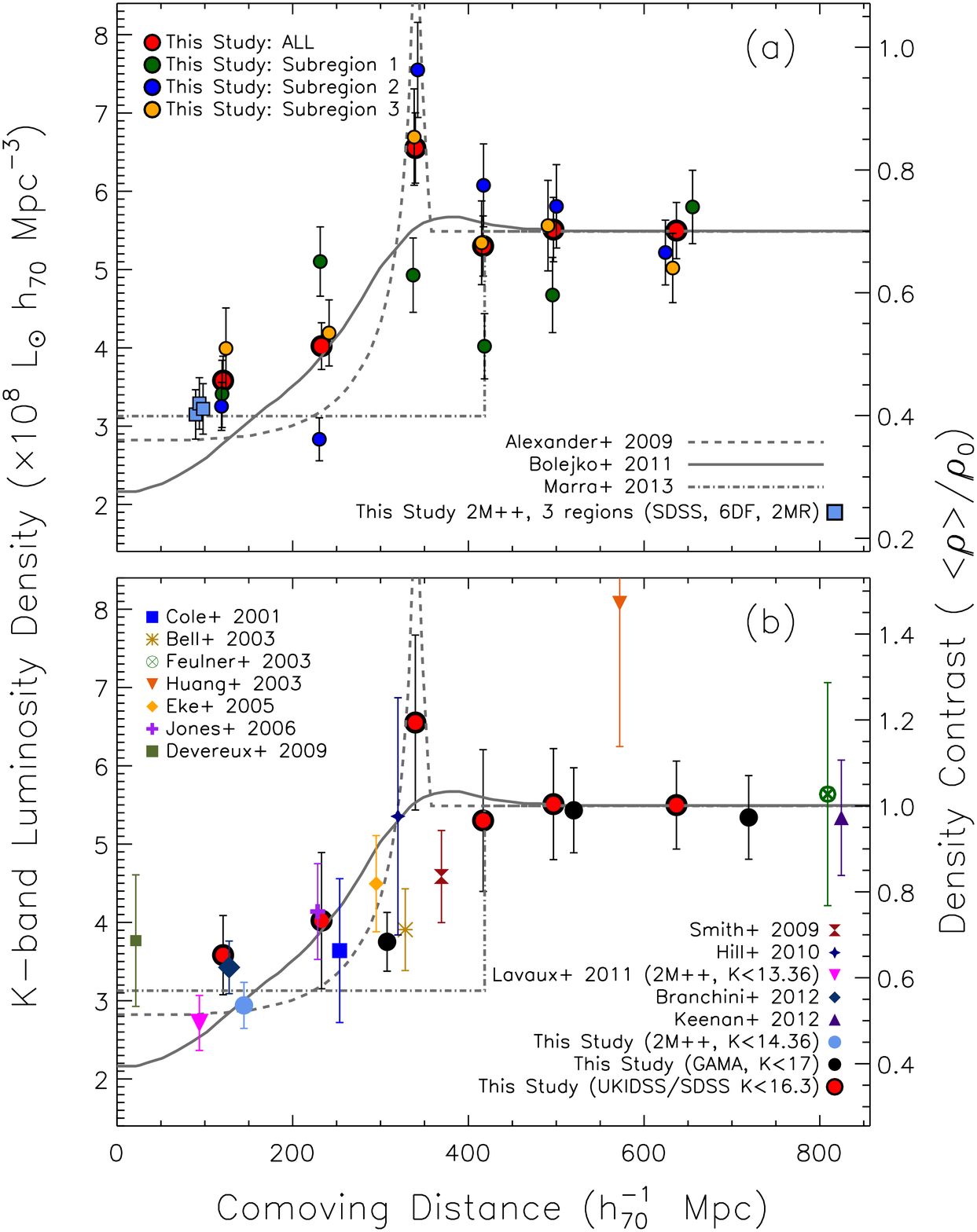}
\caption{\label{ukld} $K-$band luminosity density as a function of comoving distance.  (a) Our measured $K-$band luminosity density for the full sample (red circles) versus different directions on the sky (green, blue, and orange circles corresponding to different subregions in the UKIDSS sample). Light blue squares indicate the $K-$band luminosity density we measure in three different directions (SDSS, 6DFGRS, and 2MR regions) using the 2M++ all sky catalog compiled by \citet{Lava11}.    (b) Our measured $K-$band luminosity density for the full sample (red circles) as a function of comoving distance compared with other studies from the literature.   Our estimate of the $K_{\rm{s,AB}}<14.36$ luminosity density from the 2M++ catalog (SDSS and 6DFGRS regions only) is shown as a light blue circle.  Our estimates in three redshift bins for the GAMA survey only ($K_{\rm{AB}}<17$, same methods as for the UKIDSS sample) are shown as black circles.  The density contrast, $\langle \rho \rangle / \rho_0$, is displayed on the right-hand vertical axis.  The scale of the right-hand axis was established by performing an error-weighted least-squares fit (for the normalization only, not shape) of the radial density profile of \citet{Bole11a} (gray solid curve) to all the luminosity density data in panel (b).  The dashed curve shows the radial density profile of \citet{Alex09}.  Both \citet{Alex09} and \citet{Bole11a} claimed these density profiles can provide for good fits to the SNIa data without dark energy.   The dash-dot curve shows the scale and amplitude of the ``Hubble bubble" type perturbation that \citet{Marr13} would require to explain the discrepancy between local measurements of the Hubble constant and those inferred by Planck.}
\end{center}
\end{figure}

\end{document}